\documentclass[%
 print,
 amsmath,amssymb,
aps,
]{revtex4-2}

\usepackage{graphicx}% Include figure files
\usepackage{dcolumn}% Align table columns on decimal point
\usepackage{bm}% bold math
\usepackage{multirow}
\usepackage{tabularx}
\renewcommand{\d}[1]{\frac{\mathrm{d} #1}{\mathrm{d} t}}
\renewcommand{\b}[1]{\boldsymbol{#1}}

\begin{document}

\title{From resolvent to Gramians:\\extracting forcing and response modes for control}
% \thanks{footnote}%

\author{Benjamin Herrmann}
\affiliation{Department of Mechanical Engineering, University of Chile, Beauchef 851, Santiago, Chile}
\email{benjaminh@uchile.cl}

\author{Peter J. Baddoo}
\affiliation{Department of Mathematics, Massachusetts Institute of Technology, Cambridge, MA 02139, USA}

\author{Scott T. M. Dawson}
\affiliation{Mechanical, Materials, and Aerospace Engineering Department, Illinois Institute of Technology, Chicago, IL 60616, USA}

\author{Richard Semaan}
\affiliation{Institute of Fluid Mechanics, Technische Universit\"at Braunschweig, 38108 Braunschweig, Germany}

\author{Steven L. Brunton}
\affiliation{Department of Mechanical Engineering, University of Washington, Seattle, WA 98195, USA}

\author{Beverley J. McKeon}
\affiliation{Department of Mechanical Engineering, Stanford University, Stanford, CA 94305, USA}

% \date{\today}% It is always \today, today,
             %  but any date may be explicitly specified

\begin{abstract}
During the last decade, forcing and response modes produced by resolvent analysis have demonstrated great potential to guide sensor and actuator placement and design in flow control applications.
However, resolvent modes are frequency-dependent, which, although responsible for their success in identifying scale interactions in turbulence, complicates their use for control purposes.
In this work, we seek orthogonal bases of forcing and response modes that are the most responsive and receptive, respectively, across all frequencies.
We show that these frequency-independent bases of \emph{representative} resolvent modes are given by the eigenvectors of the observability and controllability Gramians of the system considering full state inputs and outputs. 
We present several numerical examples where we leverage these bases by building orthogonal or interpolatory projectors onto the dominant forcing and response subspaces.
Gramian-based forcing modes are used to identify dynamically relevant disturbances, to place point sensors to measure disturbances, and to design actuators for feedforward control in the subcritical linearized Ginzburg--Landau equation.
Gramian-based response modes are used to identify coherent structures and for point sensor placement aiming at state reconstruction in the turbulent flow in a minimal channel at $\mathrm{Re}_{\tau}=185$.
The approach does not require data snapshots and relies only on knowledge of the steady or mean flow.
%Max length 500 words.
\end{abstract}

%\keywords{Suggested keywords}%Use showkeys class option if keyword
                              %display desired
\maketitle

%\tableofcontents

\section{\label{sec:intro}Introduction}

Advances in our ability to control complex fluid flows to manipulate aerodynamic forces, reduce noise, or promote mixing, have a profound impact on the performance of engineering systems found in aeronautics, energy-conversion, and transportation, among others \citep{gadelhak1989arfm}. However, industrially relevant flows are typically high-dimensional and nonlinear dynamical systems, which makes their control very challenging~\citep{brunton2015amr}. Fortunately, the dynamics of these systems is often dominated by a few physically meaningful flow structures, or modes, that can be extracted from the governing equations or learned from data using modal decompositions~\citep{taira2017aiaa,taira2019aiaa}. Moreover, powerful tools from linear control theory can be leveraged if the system of interest is amenable to linearization about a steady state~\citep{bagheri2009amr,sipp2010amr,fabbiane2014amr,sipp2016amr}.\\

Resolvent analysis is a modal decomposition, based on governing equations in the frequency-domain, that is particularly useful to characterize non-normal systems, such as shear and advection-dominated flows~\citep{trefethen1993science,schmid2007arfm}. The resolvent operator governs how any harmonic forcing, at a specific frequency, is amplified by the dynamics to produce a response. Its decomposition produces resolvent gains and resolvent forcing and response modes that enable low-rank approximations of the forcing-response dynamics of the full system, which are extremely valuable for modeling, controlling, and understanding the underlying flow physics~\citep{mckeon2017jfm,jovanovic2021arfm}. Although resolvent analysis was initially conceived for laminar flows linearized about a steady flow~\citep{trefethen1993science}, \citet{mckeon2010jfm} showed that the analysis of mean-flow-linearized turbulent flows is able to uncover scale interactions and provide insight into self-sustaining mechanisms. Data-driven resolvent analysis is a recent method, developed by \citet{herrmann2021jfm}, to compute resolvent modes and gains directly from time-resolved measurement data, and without requiring the governing equations, by leveraging the dynamic mode decomposition (DMD)~\citep{schmid2010jfm,rowley2009jfm} or any of its variants~\cite{tu2014jcd,schmid2022arfm,baddoo2021arxiv}. It is important to remark that the data-driven version was developed for linear systems, therefore, its application to turbulent flows requires separating the nonlinear contributions to the dynamics from the data using, for example, the linear and nonlinear disambiguation optimization (LANDO)~\citep{baddoo2022prsa}.\\

During the last decade, resolvent analysis has been applied to several control-oriented tasks, including reduced-order modeling, data assimilation, estimation, sensor and actuator placement, and open- and closed-loop control. Resolvent-based reduced-order models have been studied for turbulent channel flows~\citep{moarref2013jfm,moarref2014pof,mckeon2017jfm,abreu2020ijhff}, laminar and turbulent cavity flows~\citep{gomez2016jfm,sun2020aiaa}, and turbulent jets~\citep{schmidt2018jfm,lesshafft2019prf}. \citet{symon2020aiaa} and \citet{franceschini2021arxiv} developed data-assimilation frameworks that leverage resolvent analysis to reconstruct mean and unsteady flow fields from a few experimental measurements. \citet{towne2020jfm} formulated a resolvent-based method to estimate space-time flow statistics from limited data. Subsequently, \citet{martini2020jfm} and \citet{amaral2021jfm} developed and applied, respectively, an optimal and non-causal resolvent-based estimator that is able to reconstruct unmeasured, time-varying, flow quantities from limited experimental data as post-processing.\\

In the context of control, a very important landmark is the input--output framework adopted by~\citet{jovanovic2005jfm} that allows one to focus on the response of certain outputs of interest, such as sparse sensor measurements, to forcing of specific input components, such as localized actuators. This allows  investigating the responsivity and receptivity of forcings and responses restricted to subsets of inputs and outputs generated by feasible actuator and sensor configurations, respectively. Therefore, the analysis provides insight into the frequencies and spatial footprints of harmonic forcings that are efficient at modifying the flow and the responses they generate. The simplest approach considers placing actuators and sensors so that they align with the forcing and response modes found and exciting their corresponding frequencies. However, although these actuators can efficiently modify the flow when used at their specific frequencies, the method is not able to determine the sense of this change with respect to a control objective, and provides no guarantees regarding the control authority over a range of frequencies. Nevertheless, this approach has been extensively used to provide guidelines for sensor and actuator placement and control strategies. In this way, resolvent analysis has been successfully leveraged to delay transition in a flat-plate boundary layer~\citep{bagheri2009jfm}, control the onset of turbulence and reduce turbulent drag in a channel flow via streamwise traveling waves~\citep{moarref2010jfm} and transverse wall oscillations~\citep{moarref2012jfm}, respectively, mitigate the effect of stochastic disturbances in the laminar flow behind a backward-facing step~\citep{boujo2015jfm}, predict noise generation in a turbulent axisymmetric jet~\citep{jeun2016pof}, understand transition mechanisms in a parallel-disk turbine~\cite{herrmann2018am}, promote mixing in a liquid-cooled heat sink \citep{herrmann2018hmt}, control separation in the turbulent flow over a NACA 0012 airfoil \citep{yeh2019jfm}, suppress oscillations in a supersonic turbulent cavity flow \citep{liu2021jfm_b}, and reduce the size of a turbulent separation bubble using  zero-net-mass-flux actuation~\citep{wu2022prf}. Recently,~\citet{skene2022jfm} developed an optimization framework to find sparse resolvent forcing modes, with small spatial footprints, that can be targeted by more realistic localized actuators. Similarly,~\citet{lopez2023scitech} introduced a sparsity-promoting resolvent analysis that allows identification of responsive forcings that are localized in space and time. Another important advancement for this framework is the structured input--output analysis, developed by~\citet{liu2021jfm}, that preserves certain properties of the nonlinear forcing and is able to recover transitional flow features that were previously available only through nonlinear input--output analysis~\citep{rigas2021jfm}. \\

The works of~\citet{luhar2014jfm} and \citet{toedtli2019prf} used resolvent analysis to study the effect of opposition control for drag reduction in wall-bounded turbulent flows. Their approach incorporates feedback control directly into the resolvent operator via the implementation of a boundary condition that accounts for wall blowing/suction with a strength proportional to sensor readings at a fixed height over the wall. The work of~\citet{leclercq2019jfm} was the first to account for mean flow deformation due to control, iterating between resolvent-guided controller design and computing the resulting controlled mean flow. Another promising approach is that of \citet{martini2022jfm}, that introduced the Wiener-Hopf formalism to perform resolvent-based optimal estimation and control of globally stable systems with arbitrary disturbance statistics. The work of~\citet{jin2022tcfd}, to our knowledge, is the first to leverage resolvent analysis to directly address the challenge of flow control with high-dimensional inputs and outputs. They develop a technique to reduce the number of inputs and outputs by projecting onto orthogonal bases for their most responsive and receptive components over a range of frequencies, which they refer to as terminal reduction. The method is then applied for optimal control and estimation of the cylinder flow at low Reynolds numbers.\\

Resolvent modes are frequency-dependent flow structures that can be physically interpreted as the most responsive forcings and the most receptive responses at a given frequency. However, for certain control-oriented tasks, such as terminal reduction~\citep{jin2022tcfd}, it is preferable to have modes that are meaningful over all (or a range of) frequencies. For example, the proper orthogonal decomposition (POD)~\citep{lumleybook} produces frequency-agnostic modes and has been leveraged to identify coherent structures in turbulent flows~\citep{berkooz1993arfm}, to build reduced-order models via Galerkin projection~\citep{rowley2017arfm}, and for sparse sensor placement for reconstruction~\citep{manohar2018csm}. Similarly, balancing modes, originally developed for model reduction in the seminal work of~\citet{moore1981ieeetac}, have been leveraged for  system identification from direct and adjoint data snapshots, both in the time-~\citep{willcox2002aiaa,rowley2005ijbc} and frequency-domains~\citep{zhou1999ijrnc,dergham2011pof}, and sparse sensor and actuator placement for feedback control~\citep{manohar2021tac}. Balancing transformations are also intimately connected to the eigensystem realization algorithm (ERA)~\citep{juang1985jgcd}, an input--output system identification technique that has been widely successful in a range of linear flow control applications~\citep{cabell2006aiaa,ahuja2010jfm,illingworth2012jfm,belson2013pof,brunton2013jfm,brunton2014jfs,illingworth2016tcfd,flinois2016jfm}.\\

\begin{figure}[t]
    \centering
    \includegraphics[width=1\textwidth]{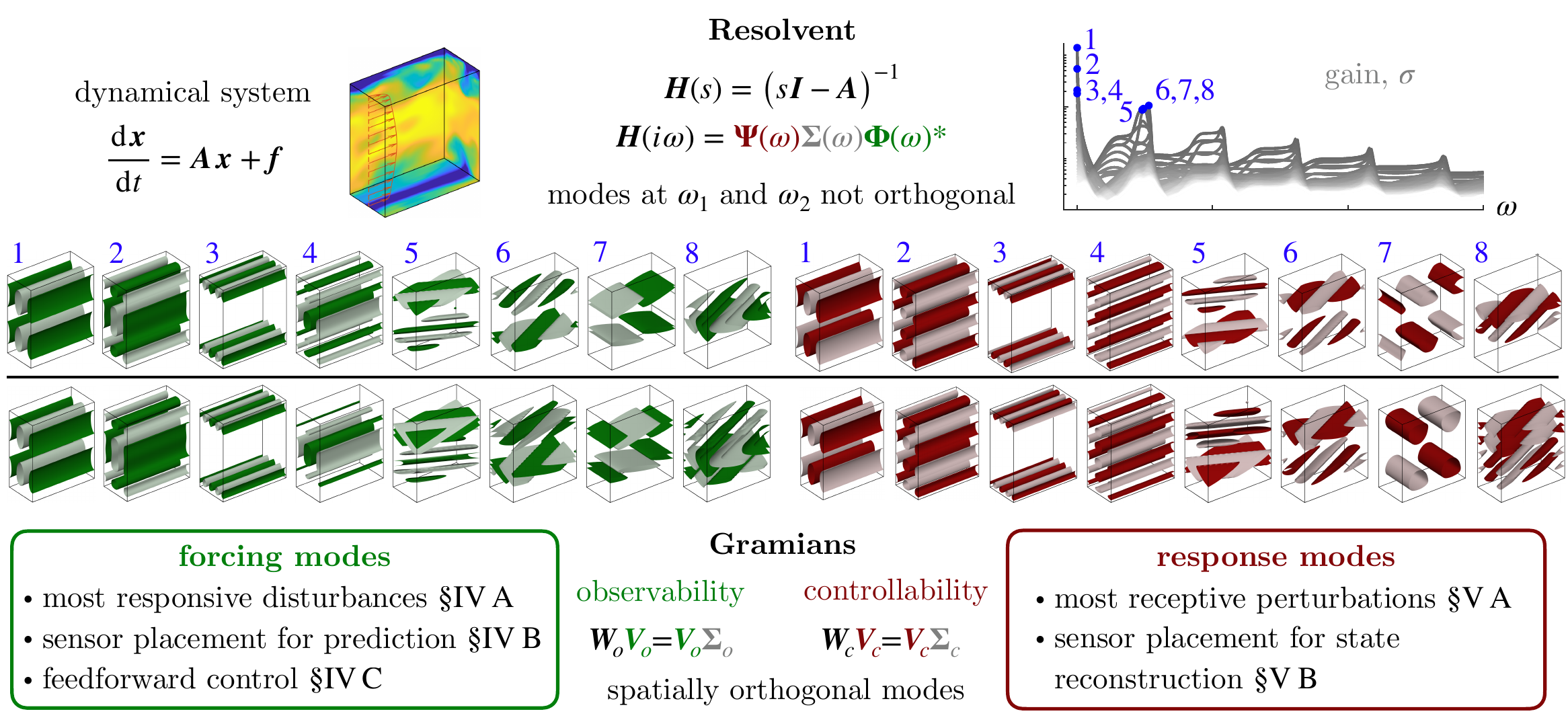}
    \caption{Forcing and response resolvent modes at handpicked frequencies (top) compared to the leading eigenvectors of the observability and controllability Gramians (bottom) for mean-flow-linearized minimal channel flow at $\mathrm{Re}_\tau=185$.
    Gramian eigenmodes span the same forcing or response subspace as the set of the corresponding resolvent modes at all frequencies, trading frequency-content information for spatial orthogonality, thus making them better suited for the control-oriented applications indicated. Real part of forcing and response modes depicted as isosurfaces of wall-normal forcing and streamwise velocity, respectively.}
    \label{fig:resolvent2gramians}
\end{figure}

In this work, we find that the notions of \emph{responsivity} and \emph{receptivity} tied to resolvent forcing and response modes, respectively, are connected to the concepts of \emph{observability} and \emph{controllability} from control theory. We show that the eigenvectors of the observability and controllability Gramians, defined in \S\ref{sec:main1}, can be interpreted as forcing and response modes in a similar fashion as resolvent modes can. In fact, forcing and response modes produced by the Gramians span the same subspaces as the sets of forcing and response resolvent modes at all frequencies. However, the Gramian eigenmodes lose the frequency-content information in favor of orthogonality with respect to a spatial inner-product, as shown in Fig.~\ref{fig:resolvent2gramians}. This trade-off makes the Gramian eigenmodes better suited to produce low-rank approximations of the dominant forcing and response subspaces across all frequencies, thus facilitating their use for control oriented applications.\\

The remainder of the paper is organized as follows. 
The mathematical connection between resolvent modes and Gramian eigenmodes is derived in section \S\ref{sec:main}. The idea of forcing and response projections for control related tasks is explained in \S\ref{sec:methods}. Examples leveraging the Gramian-based forcing and response modes are presented in \S\ref{sec:forcing} and \S\ref{sec:response}, respectively. Our conclusions are offered in \S\ref{sec:conclusions}.

\section{From resolvent to Gramians\label{sec:main}}

In this section, we begin with a brief background on resolvent analysis and on the Gramians—operators arising in linear control theory. This is followed by the new mathematical connections drawn between resolvent forcing and response modes and the eigenvectors of the observability and controllability Gramians. Finally, we justify the interpretation of the latter as forcing and response modes that are representative across all frequencies.

\subsection{Background: the resolvent operator}

Let us consider a forced linear dynamical system
\begin{equation}
\d{\b{x}}=\b{A}\b{x}+\b{f}, \label{sys}
\end{equation}
where $t\in\mathbb{R}$ denotes time, $\b{x}(t)\in \mathbb{R}^n$ is the state whose dynamics are governed by the operator $\b{A}\in \mathbb{R}^{n\times n}$, and $\b{f}\in \mathbb{R}^n$ is the forcing. Such a system may arise from a semi-discretized partial differential equation, and in the case of fluid flows, the incompressible Navier--Stokes equations can be written in this form by projecting the velocity field onto a divergence-free basis to eliminate the pressure variable. The state $\b{x}$ may either represent the deviation from a steady state of a laminar flow, or fluctuations about the temporal mean of a statistically stationary unsteady flow. In both cases the matrix $\b{A}$ is the linearization of the underlying nonlinear system about the corresponding base flow, either the equilibrium or mean flow. The forcing term may represent disturbances from the environment, model discrepancy, open-loop control actuation and/or the effect of nonlinear terms. Although we have defined $\b{A}$, $\b{x}$ and $\b{f}$ to be real for simplicity, the methods presented below can be equally applied in the case of complex variables.\\

In the Laplace $s$-domain, the resolvent operator $\b{H}(s)=(s\b{I}-\b{A})^{-1}$ is the transfer function from all possible inputs to all possible outputs in state space, with $\b{I}\in \mathbb{R}^{n\times n}$ being the identity matrix. This operator encodes how any harmonic forcing $\hat{\b{f}}(\omega)e^{i\omega t}$ at a specific frequency $\omega$ is amplified by the linear dynamics to produce a response
\begin{equation}
\hat{\b{x}}(\omega)=\b{H}(i\omega)\hat{\b{f}}(\omega).
\end{equation}

At any given frequency, a singular value decomposition (SVD) of the resolvent operator reveals the most responsive forcings, their gains, and the most receptive responses at that particular frequency. Specifically, the SVD factorizes the resolvent into $\b{H}(i\omega)=\b{\Psi}(\omega)\b{\Sigma}(\omega)\b{\Phi}(\omega)^*$, where $\b{\Sigma} \in \mathbb{R}^{n\times n}$ is a diagonal matrix containing the resolvent gains $\sigma_1 \geq \sigma_2 \geq \dotsc\geq \sigma_n \geq 0$, and $\b{\Phi}= \ [\b{\phi}_1 \ \b{\phi}_2\ \dotsc \ \b{\phi}_n]\in \mathbb{C}^{n\times n}$ and $\b{\Psi}=[\b{\psi}_1 \ \b{\psi}_2\ \dotsc \ \b{\psi}_n]\in \mathbb{C}^{n\times n}$ are unitary matrices whose columns are known as the forcing and response resolvent modes, $\b{\phi}_j$ and $\b{\psi}_j$, respectively.\\

\subsection{Background: the Gramians}

Linear control theory deals with systems of the form
\begin{equation}
    \d{\b{x}}=\b{Ax}+\b{Bu}, \qquad \b{y} = \b{Cx},
\end{equation}
where the matrices $\b{B} \in \mathbb{R}^{n\times p}$ and $\b{C}\in \mathbb{R}^{q \times n}$ are determined by the configuration of actuators and sensors, and describe how the control inputs $\b{u}(t)\in \mathbb{R}^p$ act on the dynamics and which outputs $\b{y}(t)\in \mathbb{R}^q$ are measured from the state, respectively. Note that $\b{B}$, $\b{C}$, $\b{u}$ and $\b{y}$ might be complex as well. In this setting, it is often important to identify the most \emph{controllable} and \emph{observable} directions in state-space, that is, what are the states that are most efficiently excited by the inputs, and what are the states that are most easily discerned from the measured outputs. This is achieved by analyzing the infinite-time observability and controllability Gramians, $\b{W_o}$ and $\b{W_c}$, defined as
\begin{subequations}
\begin{align}
    \b{W_o} &= \int_{0}^\infty e^{\b{A}^*t}\b{C}^*\b{C}e^{\b{A}t} ~\mathrm{d}t = \frac{1}{2\pi}\int_{-\infty}^\infty \b{H}(i\omega)^*\b{C}^*\b{C}\b{H}(i\omega) ~\mathrm{d}\omega, \\
    \b{W_c} &= \int_{0}^\infty e^{\b{A}t}\b{B}\b{B}^*e^{\b{A}^*t} ~\mathrm{d}t = \frac{1}{2\pi}\int_{-\infty}^\infty \b{H}(i\omega)\b{B}\b{B}^*\b{H}(i\omega)^* ~\mathrm{d}\omega,
\end{align}
\end{subequations}
where the rightmost expressions are the frequency-domain representations which, as is well-known, feature the resolvent operator. It is important to point out that these representations are equivalent for stable systems only; whereas the integrals in the time-domain diverge for the unstable case, the definitions in the frequency-domain still hold as long as there are no marginal eigenvalues~\citep{zhou1999ijrnc,dergham2011pof}. Both matrices, $\b{W_o}$ and $\b{W_c}$, are symmetric positive semi-definite, and their eigenvectors comprise hierarchically-ordered orthogonal bases of the most observable and controllable states, respectively.\\

\subsection{Eigenvectors of the Gramians are representative resolvent modes\label{sec:main1}}

We can now go back to a system of the form in Eq.~\eqref{sys} by considering $\b{C}=\b{B}=\b{I}$ such that we are measuring all possible outputs $\b{y}=\b{x}$ and forcing all possible inputs $\b{u}=\b{f}$. In this scenario, the Gramians become
\begin{subequations}
\begin{align}
    \b{W_o} &= \int_{0}^\infty e^{\b{A}^*t}e^{\b{A}t} ~\mathrm{d}t = \frac{1}{2\pi}\int_{-\infty}^\infty \b{H}(i\omega)^*\b{H}(i\omega) ~\mathrm{d}\omega, \\
    \b{W_c} &= \int_{0}^\infty e^{\b{A}t}e^{\b{A}^*t} ~\mathrm{d}t = \frac{1}{2\pi}\int_{-\infty}^\infty \b{H}(i\omega)\b{H}(i\omega)^* ~\mathrm{d}\omega.
\end{align}
\end{subequations}
Furthermore, substituting the resolvent by its singular value decomposition $\b{H}(i\omega)=\b{\Psi}(\omega)\b{\Sigma}(\omega)\b{\Psi}(\omega)^*$ and using the orthogonality of the resolvent modes at each $\omega$ yields
\begin{equation}
    \b{W_o} = \frac{1}{2\pi}\int_{-\infty}^\infty \b{\Phi}(\omega)\b{\Sigma}(\omega)^2\b{\Phi}(\omega)^* ~\mathrm{d}\omega, \qquad
    \b{W_c} = \frac{1}{2\pi}\int_{-\infty}^\infty \b{\Psi}(\omega)\b{\Sigma}(\omega)^2\b{\Psi}(\omega)^* ~\mathrm{d}\omega,
\end{equation}
where no approximations have been made up to this point. We now consider a numerical quadrature to represent the integrals over the frequency domain as a sum over $m$ discrete frequencies $\omega_k$. For ease of notation, we consider a trapezoidal rule with a uniform frequency-spacing $\Delta \omega$, but other quadratures and non-uniform spacing could be considered in general. Letting $\sigma_{jk}$, $\b{\phi}_{jk}$ and $\b{\psi}_{jk}$ denote the $j$-th resolvent gain, forcing, and response modes evaluated at the $k$-th frequency leads to the following approximations of the Gramians
\begin{subequations}
\begin{align}
    \b{W_o}  &\approx \frac{\Delta \omega}{2\pi}\sum_{k=1}^m \b{\Phi}(\omega_k)\b{\Sigma}(\omega_k)^2\b{\Phi}(\omega_k)^*\\
     &= \frac{\Delta \omega}{2\pi}\sum_{k=1}^m\left[\sigma_{1k}\b{\phi}_{1k} \ \cdots \ \sigma_{nk}\b{\phi}_{nk}\right]
     \left[\begin{array}{c}
          \sigma_{1k}\b{\phi}_{1k}^*   \\
           \vdots\\
          \sigma_{nk}\b{\phi}_{nk}^*
     \end{array}\right] \\
          &= \frac{\Delta \omega}{2\pi}\left[\sigma_{11}\b{\phi}_{11} \ \cdots \ \sigma_{nm}\b{\phi}_{nm}\right]
     \left[\begin{array}{c}
          \sigma_{11}\b{\phi}_{11}^*   \\
           \vdots\\
          \sigma_{nm}\b{\phi}_{nm}^*
     \end{array}\right] \approx \b{L_o}\b{L_o}^*,\label{Lo}
\end{align}
\end{subequations}
\begin{subequations}
\begin{align}
    \b{W_c}  &\approx \frac{\Delta \omega}{2\pi}\sum_{k=1}^m \b{\Psi}(\omega_k)\b{\Sigma}(\omega_k)^2\b{\Psi}(\omega_k)^*\\
     &= \frac{\Delta \omega}{2\pi}\sum_{k=1}^m\left[\sigma_{1k}\b{\psi}_{1k} \ \cdots \ \sigma_{nk}\b{\psi}_{nk}\right]
     \left[\begin{array}{c}
          \sigma_{1k}\b{\psi}_{1k}^*   \\
           \vdots\\
          \sigma_{nk}\b{\psi}_{nk}^*
     \end{array}\right] \\
          &= \frac{\Delta \omega}{2\pi}\left[\sigma_{11}\b{\psi}_{11} \ \cdots \ \sigma_{nm}\b{\psi}_{nm}\right]
     \left[\begin{array}{c}
          \sigma_{11}\b{\psi}_{11}^*   \\
           \vdots\\
          \sigma_{nm}\b{\psi}_{nm}^*
     \end{array}\right] \approx \b{L_c}\b{L_c}^*,\label{Lc}
\end{align}
\end{subequations}
where the expressions in Eqs.~\eqref{Lo} and \eqref{Lc} reorder terms as the multiplication of two matrices, allowing us to recognize the Cholesky factors of the Gramians $\b{L_o}\approx \left(\Delta \omega/2\pi\right)^{1/2}\left[\sigma_{11}\b{\phi}_{11} \ \cdots \ \sigma_{nm}\b{\phi}_{nm}\right]$ and $\b{L_c}\approx \left(\Delta \omega/2\pi\right)^{1/2}\left[\sigma_{11}\b{\psi}_{11} \ \cdots \ \sigma_{nm}\b{\psi}_{nm}\right]$. It is now clear that the column space of these Cholesky factors span the same forcing or response subspace as all the corresponding resolvent modes at all frequencies. Moreover, the left singular vectors of $\b{L_o}$ and $\b{L_c}$, which are precisely the eigenvectors of $\b{W_o}$ and $\b{W_c}$, form orthogonal bases for these forcing and response subspaces. Next, we justify the physical interpretation of these eigenvectors as forcing and response modes.

\subsection{Gramian-based forcing and response modes}

We seek the most responsive forcing and the most receptive response across all frequencies. This endeavor can be formalized as the optimization problems presented below.\\

First, let us introduce the $\mathcal{L}_2-$norm for signals $\b{v}(t)$, which measures their energy integrated over all time, as follows
\begin{equation}
    \|\b{v}(t)\|_{\mathcal{L}_2}^2=\int_0^\infty \|\b{v}(t)\|_2^2 ~\mathrm{d}t=\frac{1}{2\pi}\int_{-\infty}^\infty \|\hat{\b{v}}(\omega)\|_2^2~\mathrm{d}\omega 
    =\|\hat{\b{v}}(\omega)\|_{\mathcal{L}_2}^2,
\end{equation}
where the frequency-domain expression follows from Plancherel's theorem. Therefore, in the $\mathcal{L}_2-$sense, optimizing energy amplification integrated over all frequencies is equivalent to optimizing energy growth integrated over all time.\\

Hence, the most responsive forcing across all frequencies is the mode $\b{v}$ that mapped through the resolvent produces the most amplified response in the $\mathcal{L}_2-$sense and is found by maximizing the gain
\begin{equation}
\frac{\|\b{H}(i\omega)\b{v} \|^2_{\mathcal{L}_2}}{\|\b{v}\|_2^2}
=
\frac{\|e^{\b{A}t}\b{v} \|^2_{\mathcal{L}_2}}{\|\b{v}\|_2^2}
= 
\frac{\b{v}^* \int_0^\infty e^{\b{A}^*t} e^{\b{A}t} ~\mathrm{d}t ~\b{v}}{\b{v}^*\b{v}}
= 
\frac{\b{v}^*\b{W_o} \b{v}}{\b{v}^*\b{v}},
\end{equation}
where the observability Gramian $\b{W_o}$ emerges naturally from expanding the norm definition. In fact, the cost function in the rightmost expression corresponds to a Rayleigh quotient, thus the optimizer is the leading eigenvector of $\b{W_o}$. Subsequent eigenvectors then solve for the next most responsive forcings.\\

Now, the most receptive response across all frequencies is the mode $\b{v}$ that best aligns, in the $\mathcal{L}_2-$sense, with all possible responses, which are spanned by the columns of the resolvent, thus maximizing the gain
\begin{equation}
\frac{\|\b{H}(i\omega)^*\b{v} \|^2_{\mathcal{L}_2}}{\|\b{v}\|_2^2}
=
\frac{\|e^{\b{A}^*t}\b{v} \|^2_{\mathcal{L}_2}}{\|\b{v}\|_2^2}
= 
\frac{\b{v}^* \int_0^\infty e^{\b{A}t} e^{\b{A}^*t} ~\mathrm{d}t ~\b{v}}{\b{v}^*\b{v}}
= 
\frac{\b{v}^*\b{W_c} \b{v}}{\b{v}^*\b{v}},
\end{equation}
where now a Rayleigh quotient with the controllability Gramian $\b{W_c}$ appears in the last expression. Thus, the optimizer corresponds to its leading eigenvector. Again, subsequent eigenvectors then solve for the next most receptive responses in the same sense.\\

\begin{table}[!]
    \centering
    \caption{Forcing and response modes for non-normal systems and the optimization problems they solve.}
    \begin{tabular}{m{0.37\columnwidth} m{0.2\columnwidth} m{0.30\columnwidth}}
    \hline\hline
        \multirow{2}{0.37\columnwidth}{\centering{\textbf{physical interpretation}}} & \centering{\textbf{cost maximized}} & \multirow{2}{0.25\columnwidth}{\centering{\textbf{optimizer}}}\\
         & \centering{(s.t. $\|\b{v}\|_2=1$)} & \\
        \hline\hline
        most responsive impulse forcing after $t$ & \centering{$\|e^{\b{A}t}\b{v}\|_2^2$} & leading right singular vectors of $e^{\b{A}t}$\\
        \hline
        most responsive harmonic forcing at $\omega$ & \centering{$\|\b{H}(i\omega)\b{v}\|_2^2$} & leading right singular vectors of $\b{H}(i\omega)$\\
        \hline
        most responsive impulse forcing over all $t$ & \centering{$\|e^{\b{A}t}\b{v}\|^2_{\mathcal{L}_2}$} & \multirow{2}{0.3\columnwidth}{least damped eigenvectors of $\b{W_o}$}\\
        most responsive harmonic forcing across all $\omega$ & \centering{$\|\b{H}(i\omega)\b{v}\|^2_{\mathcal{L}_2}$} & \\
        \hline
        most receptive impulse response after $t$ & \centering{$\|e^{\b{A}^*t}\b{v}\|_2^2$} & leading left singular vectors of $e^{\b{A}t}$\\
        \hline
        most receptive harmonic response at $\omega$ & \centering{$\|\b{H}(i\omega)^*\b{v}\|^2_2$} & leading left singular vectors of $\b{H}(i\omega)$\\
        \hline
        most receptive impulse response over all $t$ & \centering{$\|e^{\b{A}^*t}\b{v}\|^2_{\mathcal{L}_2}$} & \multirow{2}{0.3\columnwidth}{least damped eigenvectors of $\b{W_c}$}\\        most receptive harmonic response across all $\omega$ & \centering{$\|\b{H}(i\omega)^*\b{v}\|^2_{\mathcal{L}_2}$} & \\
        \hline
    \end{tabular}
    \label{tab:modes}
\end{table}

When working with modal decompositions of fluid flows (or dynamical systems in general), it is useful to think about the optimization problems being solved to provide physical interpretations for the modes. This has been particularly important for non-normal systems, where nonmodal stability analyses have provided valuable insights regarding physical mechanisms for linear energy amplification \cite{schmid2007arfm}. As we have shown, and summarize in Table~\ref{tab:modes}, the forcing and response modes produced by the Gramian eigenvectors are closely connected to those obtained from transient growth and resolvent analyses.\\

It is important to note that we have been using the Euclidean $2-$norm to represent the energy of a state vector $\|\b{v}\|_2^2$. It is often the case that a physically meaningful inner product considers a positive-definite weighting matrix $\b{Q}$ so that the relevant metric is $\|\b{v}\|_{\b{Q}}^2=\b{v}^*\b{Qv}$. This weighting might account for integration quadratures or scaling of heterogeneous variables in multi-physics systems~\citep{herrmann2018hmt}. However, in this scenario, one can conveniently and easily modify the problem to work with the Euclidean $2-$norm by taking the state as $\b{Fv}$ and the dynamics as $\b{F A F}^{-1}$, where $\b{F}^*$ is the Cholesky factor of $\b{Q}=\b{F}^*\b{F}$.

% \subsection{Bases of representative resolvent modes}

% Our goal is to find bases for the most responsive forcings and the most receptive responses for a linear dynamical system.

% For the case of inputs

% We wish to find a basis $V$ to project high-dimensional inputs $u(t)$ onto a lower dimensional subspace, such that the response of the system to $u(t)$ and to $u_r(t)=VV^*u(t)$ are close in some sense.

% \begin{align}
%   \|y(t)-y_r(t)\|_{\mathcal{L}_2}^2 
%   &= \frac{1}{2\pi}\int_{-\infty}^\infty \|\hat{y}(i \omega)-\hat{y}_r(i\omega)\|_2^2~\mathrm{d}\omega \\
%   &= \frac{1}{2\pi}\int_{-\infty}^\infty \|G(i\omega)\hat{u}(i \omega)-G(i\omega)\hat{u}_r(i\omega)\|_2^2~\mathrm{d}\omega
%   \\
%     &= \frac{1}{2\pi}\int_{-\infty}^\infty \|G(i\omega)(I-VV^*)\hat{u}(i \omega)\|_2^2~\mathrm{d}\omega
%     \\
%     &\le \|G(i\omega)(I-VV^*)\|_{\mathcal{H}_\infty}^2~ \|u(t)\|_{\mathcal{L}_2}^2.
% \end{align}

\section{Forcing and response projections\label{sec:methods}}

In this section we explain how and to what end one may build projectors $\mathbb{P}$ that produce low-rank approximations of forcings and responses in dynamical systems of the form in Eq.~\ref{sys}. In the case of a forcing projection, we want the response of the system $\b{x}(t)$ generated by the forcing $\b{f}(t)$ and the response $\tilde{\b{x}}(t)$ generated by its projection $\mathbb{P}\b{f}(t)$ to be \emph{close}. Specifically, we choose the $\mathcal{L}_2-$norm to quantify the error induced by the projection

\begin{subequations}\label{eq:forcing_bound}
\begin{align}
  \|\b{x}(t)-\tilde{\b{x}}(t)\|_{\mathcal{L}_2}^2 
  &= \frac{1}{2\pi}\int_{-\infty}^\infty \|\b{H}(i\omega)\hat{\b{f}}(\omega)-\b{H}(i\omega)\mathbb{P}\hat{\b{f}}(\omega)\|_2^2~\mathrm{d}\omega
   \\
    &= \frac{1}{2\pi}\int_{-\infty}^\infty \|\b{H}(i\omega)(\b{I}-\mathbb{P})\hat{\b{f}}(\omega)\|_2^2~\mathrm{d}\omega
    \\
    &\le \|\b{H}(s)(\b{I}-\mathbb{P})\|_{\mathcal{H}_\infty}^2~ \|\b{f}(t)\|_{\mathcal{L}_2}^2,
\end{align}
\end{subequations}
where $\mathcal{H}_{\infty}$ denotes the system norm defined as the supremum of the singular values of the corresponding transfer function over all frequencies. The last expression in Eqs.~\eqref{eq:forcing_bound} shows that the error of interest is bounded by the $\mathcal{H}_{\infty}$–norm of the resolvent restricted to only admit forcings that live in the null-space of the projector.\\

In the case of a response projection, we pursue a low-rank approximation $\mathbb{P}\b{x}(t)$ of the full state $\b{x}(t)$. Therefore, the $\mathcal{L}_2-$norm projection error is given by

\begin{subequations}\label{eq:response_bound}
\begin{align}
  \|\b{x}(t)-\mathbb{P}\b{x}(t)\|_{\mathcal{L}_2}^2 
  &= \frac{1}{2\pi}\int_{-\infty}^\infty \|\b{H}(i\omega)\hat{\b{f}}(\omega)-\mathbb{P}\b{H}(i\omega)\hat{\b{f}}(\omega)\|_2^2~\mathrm{d}\omega
   \\
    &= \frac{1}{2\pi}\int_{-\infty}^\infty \|(\b{I}-\mathbb{P})\b{H}(i\omega)\hat{\b{f}}(\omega)\|_2^2~\mathrm{d}\omega
    \\
    &\le \|(\b{I}-\mathbb{P})\b{H}(s)\|_{\mathcal{H}_\infty}^2~ \|\b{f}(t)\|_{\mathcal{L}_2}^2,
\end{align}
\end{subequations}
where the error is now bounded by the $\mathcal{H}_{\infty}-$norm of the resolvent restricted to only produce responses that live in the null-space of the projector.

\subsection{Orthogonal projectors}

Given an orthonormal set of vectors $V=\left[\b{v}_{1} \ \cdots \ \b{v}_{r}\right]\in \mathbb{C}^{n\times r}$, an orthogonal projector onto the column space of $\b{V}$ is given by $\mathbb{P}_{\perp}=\b{VV}^*$. If such a rank$-r$ projector minimizes the forcing projection error, measured by Eq.~\eqref{eq:forcing_bound}, then the modes in $\b{V}$ correspond to the most dynamically relevant disturbances, that is, the response of the system can be well predicted when approximating $\b{f}(t)\approx \mathbb{P}_{\perp}\b{f}(t)= \b{VV}^*\b{f}(t)=\b{V}\b{d}(t)$, with $\b{d}(t)\in \mathbb{C}^{r}$. In addition to the physical insight regarding the most responsive forcings, these modes can be leveraged to design spatially distributed actuators. Building actuators that excite these modes, resulting in a control matrix $\b{B}=\b{V}$, would produce a controller that can act along the most responsive directions in state space.\\

Similarly, if the response projection error in Eq.~\eqref{eq:response_bound} is minimized, then $\b{V}$ spans the subspace of the most receptive perturbations. In this case the modes in $\b{V}$ allow for a low-dimensional representation of the state in terms of mode amplitudes, that is $\b{x}(t)\approx \mathbb{P}_{\perp}\b{x}(t)= \b{VV}^*\b{x}(t)=\b{V}\b{a}(t)$, with $\b{a}(t)\in \mathbb{C}^{r}$.\\

Here, we propose using the eigenvectors of the observability and controllability Gramians as bases $\b{V}$ to build forcing and response projectors, respectively. In fact, these choices are optimal for the case of stable linear systems where all forcings are considered to be equally likely (e.g., white-noise disturbances). Moreover, in that scenario, the controllability Gramian eigenmodes correspond to the proper orthogonal decomposition (POD) modes obtained from the ideal dataset —where an impulse response in every degree of freedom is recorded \cite{holmesbook}. Hence, in a control setup with full state outputs, where the measurement matrix is $\b{C}=\b{I}$, the resulting response projection is equivalent to the output projection proposed by \citet{rowley2005ijbc}.

\subsection{Interpolatory projector}

Another kind of projector that is perhaps more practically useful for control purposes is an interpolatory projector $\mathbb{P}_{:}=\b{V}\left(\b{P}^{\mathrm{T}} \b{V}\right)^{-1} \b{P}^{\mathrm{T}}$, built from our basis $\b{V}$ for an $r-$dimensional subspace and a sparse matrix $\b{P}=\left[\b{e}_{{\gamma}_1} \ \cdots \ \b{e}_{{\gamma}_r}\right] \in \mathbb{R}^{n\times r}$ containing $r$ columns of the identity. Just like with the orthogonal projector $\mathbb{P}_{\perp}=\b{VV}^{\mathrm{T}}$, when $\mathbb{P}_{:}$ is applied to a vector, the result is in the column space of $\b{V}$. In addition, for the interpolatory projector, $r$ of the entries in the projected vector will match with the corresponding entries in the original vector. These entries are known as the interpolation or sampling points and are determined by the non-zero entries in the sampling matrix $\b{P}$.\\

In this case, using the leading $r$ eigenvectors of the observability Gramian as our basis, the forcing can be approximated as $\b{f}(t)\approx \mathbb{P}_{:}\b{f}(t)=\b{V}\left(\b{P}^{\mathrm{T}} \b{V}\right)^{-1} \b{P}^{\mathrm{T}}\b{f}(t)=\b{V}\left(\b{P}^{\mathrm{T}} \b{V}\right)^{-1}\b{d}(t)$, where now $\b{d}$ corresponds to the entries of $\b{f}$ at the $r$ interpolation points. This means that we can approximate the forcing in the subspace of the most responsive disturbances from sparse sensor measurements and, if a model of the dynamics is available, predict the response of the system. Similarly, using the controllability Gramian eigenvectors in $\b{V}$, we can reconstruct the response $\b{x}(t)\approx \mathbb{P}_{:}\b{x}(t)=\b{V}\left(\b{P}^{\mathrm{T}} \b{V}\right)^{-1} \b{P}^{\mathrm{T}}\b{x}(t)=\b{V}\left(\b{P}^{\mathrm{T}} \b{V}\right)^{-1}\b{a}(t)$ from sparse sensor measurements $\b{a}$ corresponding to the entries of $\b{x}$ at the $r$ interpolation points.\\

The choice of the sensor locations, and therefore the design of $\b{P}$, is critical for the quality of the projection. In addition, when using the Gramian eigenvectors in $\b{V}$, the interpolation points that result in small projection errors are physically interesting, since they represent the most responsive or receptive spatial locations. However, finding the optimal sampling matrix that minimizes the projection error for a given basis $\b{V}$ requires a brute force search over all possible sensor location combinations. While this can be achieved for small-scale systems \citep{chen2011jfm}, with $n\sim\mathcal{O}(10^4)$ at the moment of this writing, the computational cost makes it intractable in higher dimensions. Alternatively, there are several fast greedy algorithms that can be used to approximate the optimal sensor locations avoiding the combinatorial search. In the context of reduced-order modeling, the empirical and discrete empirical interpolation methods, EIM \citep{barrault2004crm} and DEIM \citep{chaturantabut2010sisc}, were developed to find locations to interpolate nonlinear terms in a high-dimensional dynamical system, which is known as hyper-reduction. An even simpler and equally efficient approach is the Q-DEIM algorithm, introduced by \citet{drmac2016sisc}, that leverages the pivoted QR factorization to select the sampling points. \citet{manohar2018csm} showed that this is also a robust strategy for sparse-sensor placement for state reconstruction on a range of applications. Here, we use pivoted QR to design $\b{P}$ because it provides near-optimal interpolation points and is simple to implement.

\subsection{About balancing modes}

Balancing modes, introduced in the seminal work of \citet{moore1981ieeetac}, correspond to the most jointly controllable and observable states of a linear dynamical system. The adjoint and direct balancing modes are computed as the columns of $\b{L_o U S}^{-1/2}$ and $\b{L_c T S}^{-1/2}$, where $\b{L_o}$ and $\b{L_c}$ are the Cholesky factors of the observability and controllability Gramians, respectively, and the matrices $\b{U}$, $\b{S}$ and $\b{T}$ come from the SVD of $\b{L_o}^*\b{L_c}=\b{UST}^*$ truncated at some rank $r$ \citep{rowley2017arfm}. Although these modes are also an attractive choice to build forcing and response projectors, they represent sub-optimal forcings and responses in the $\mathcal{L}_2-$sense, since the balancing implies a trade-off between observability (\emph{responsivity}) and controllability (\emph{receptivity}). Therefore, for the tasks presented in this work, although balancing modes might lead to similar results with only a slight drop in performance, the eigenvectors of the Gramians are a more principled choice because of the clear connection to resolvent analysis and because they correspond to the optimal forcings and responses, as explained in \S\ref{sec:main}. Nevertheless, balanced coordinates are better suited for other control-oriented tasks. For example, balanced proper orthogonal decomposition (BPOD) uses empirical approximations of the balancing modes to perform model reduction of high-dimensional linear dynamical systems \citep{willcox2002aiaa,rowley2005ijbc}, and, more recently, the covariance balancing reduction using adjoint snapshots (CoBRAS), developed by \citet{otto2022arxiv}, is able to tackle nonlinear model reduction. Another example is the recent work of \citet{manohar2021tac} that successfully leveraged the balancing modes to inform the placement of sensors and actuators for feedback control. 

\section{Leveraging forcing modes\label{sec:forcing}}

In this section, we present three numerical examples detailing how the Gramian-based forcing modes can be leveraged to: identify the most responsive disturbances, predict the response to disturbances interpolated from sparse measurements, and perform disturbance feedforward control. Our test bed is the linearized complex Ginzburg--Landau equation, which is a typical model for instabilities in spatially-evolving flows. The semi-discretized system is governed by the linear operator
\begin{equation}
\b{A} = -\nu \b{D}_x + \gamma \b{D}_x^2 + \mu(x),\label{eq:cgl_A}
\end{equation}
where $x$ is the spatial coordinate, and $\b{D}_x$ and $\b{D}^2_x$ are the $1^{st}$ and $2^{nd}$-order spatial differentiation matrices with homogeneous boundary conditions at $x\rightarrow \pm \infty$. 
We choose a quadratic spatial dependence for the parameter $\mu(x)=(\mu_0-c_{\mu}^2)+\tfrac{\mu_2}{2}x^2$, that has been used previously by several authors \citep{bagheri2009amr,chen2011jfm,towne2018jfm}. 
The other parameters are set to $\mu_0=0.23$, $\mu_2=-0.01$, $\nu=2+0.4i$, and $\gamma=1-i$ as in \citet{towne2018jfm}, giving rise to linearly stable dynamics. 
As in \citet{bagheri2009amr}, we use spectral collocation based on Gauss-weighted Hermite polynomials to build the differentiation matrices $\b{D}_x$ and $\b{D}^2_x$ and the integration quadrature. 
The spatial coordinate is discretized into $n=220$ collocation points, and the domain is truncated to $x\in [-85,85]$, which is sufficient to enforce the far-field boundary conditions. The observability Gramian considering full state inputs and outputs is computed using readily available routines that solve the corresponding Lyapunov equation.\\

\begin{figure}[t]
    \centering
    \includegraphics[width=1\textwidth]{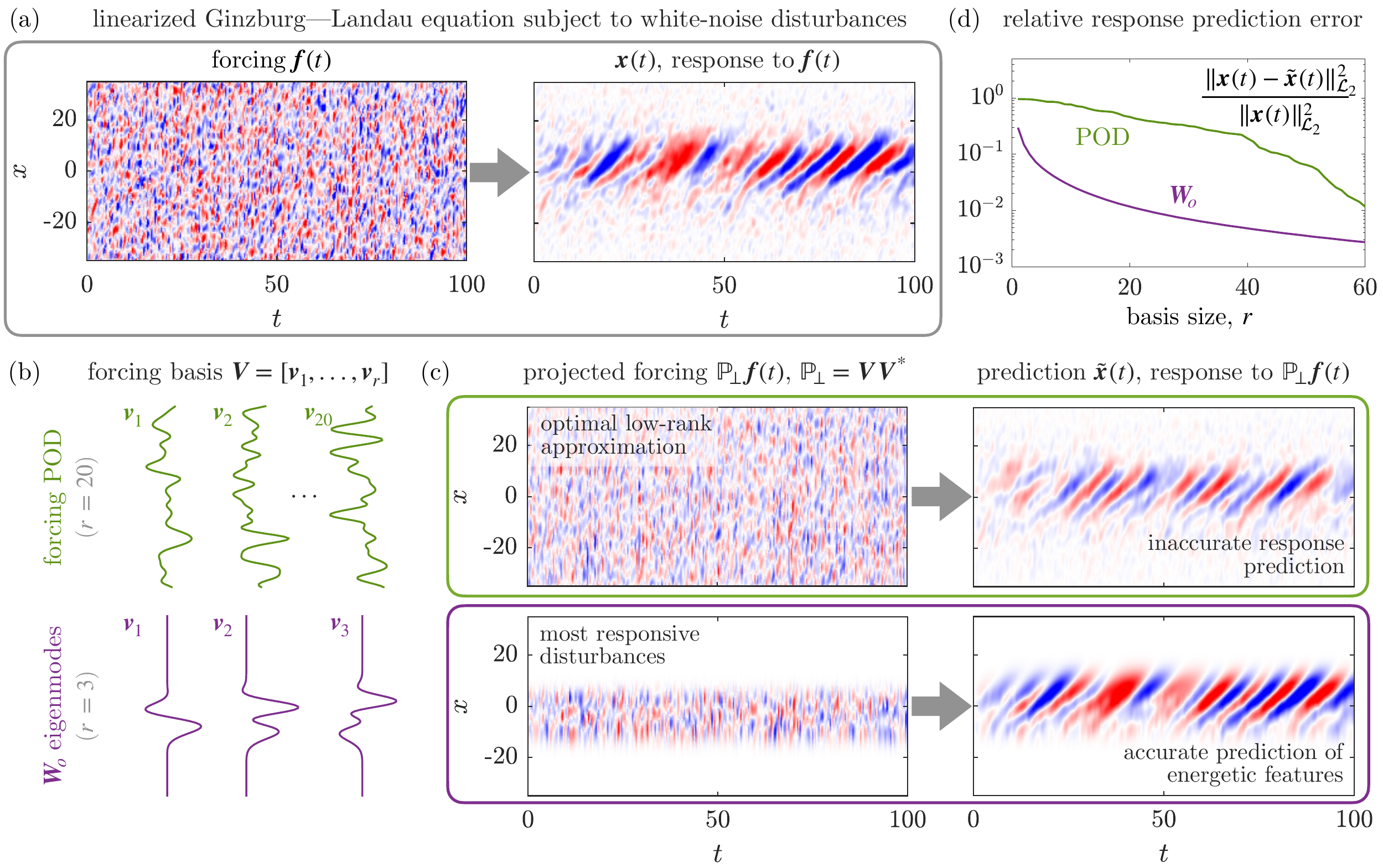}
    \caption{(a) Response of the linearized complex Ginzburg--Landau equation to white noise forcing. The system behaves as a selective disturbance amplifier. (b) Two bases of representative forcing modes are used to build orthogonal projectors: the leading forcing POD modes computed from snapshots of the disturbances, and the leading eigenvectors of the observability Gramian considering full state inputs and outputs. (c) Projected forcings and their induced responses using $20$ forcing POD modes vs $3$ $\b{W_o}$ eigenmodes. (d) Normalized $\mathcal{L}_2$-error in the response prediction using projected forcings as a function of the projector rank $r$. The real part is displayed for forcing, response and mode plots.}
    \label{fig:cgl}
\end{figure}

Throughout the examples, bandlimited white noise disturbances $\b{f}(t)$ are generated, several forcing projectors $\mathbb{P}$ are built, and the responses of the system to $\b{f}(t)$ and to the projected forcings $\mathbb{P}\b{f}(t)$ are compared. Simulations are carried out by numerically integrating for $10^4$ time units and the results are recorded every $\Delta t=0.5$ time units for post-processing. Every time step, the forcing is generated on a uniform spatial grid of $80$ points as a vector of random complex numbers with amplitudes drawn from a normal distribution and phases drawn from a uniform distribution, which is then mapped onto the non-uniform spatial grid via spline interpolation.

\subsection{Most responsive disturbances}

For the set of parameters used, the linearized complex Ginzburg--Landau equation is known to behave as a selective amplifier of disturbances \citep{bagheri2009amr}. When forced with white noise, the dynamics of the system filter disturbances and generate a response where most of the energy is contained in coherent structures, as shown in Fig.~\ref{fig:cgl}(a).
This implies that, among all possible ways to excite the system, there are a few preferred forcing patterns that are dynamically relevant and have a dominant effect on the produced response.\\

We build orthogonal projectors onto two different subspaces, spanned by the bases $\b{V}$ shown in Fig.~\ref{fig:cgl}(b), that one might consider when seeking for low-rank approximations of the forcing. For the first one, we gather snapshots of the forcing, perform POD on those snapshots, and retain the leading $r$ forcing POD modes as our basis to build a rank-$r$ projector. The second subspace is the one spanned by the leading $r$ eigenvectors of the observability Gramian. The projection of the forcing onto these subspaces and the response of the system to the projected forcings are shown in Fig.~\ref{fig:cgl}(c) for $r=20$ POD modes and $r=3$ $\b{W_o}$ eigenmodes.\\

The projected forcing using POD closely resembles the original forcing, which is expected since POD provides the optimal low-rank approximation based on empirical observations. Despite this, the response is more accurately predicted using the $\b{W_o}$-based projected forcing, even when using just $3$ modes compared to the $20$ used for the POD-based projection, as shown in Fig.~\ref{fig:cgl}(c). More quantitatively, the error in the response prediction, measured by the normalized $\mathcal{L}_2$-norm, is always lower for the same forcing projection rank $r$ and decreases much more rapidly for low-rank approximations, as shown in Fig.~\ref{fig:cgl}(d).\\  

These results are not surprising since the naive POD approach only seeks to approximate the observed disturbances and is completely agnostic to the dynamics of the system. The purpose of the comparison is to highlight the dynamical significance of the eigenvectors of $\b{W_o}$ as the most responsive disturbances, providing a subspace for forcing projection that is optimal for response prediction.

% Find a set of actuation modes $V$ to minimize

% $\|G(s)-G(s)VV^* \|_{\mathcal{H}_\infty}$

% Laminar plane channel flow.

% \vspace{10cm}

% \subsection{Sensor placement to measure high-dimensional disturbances in amplifier flows}

\begin{figure}[t]
    \centering
    \includegraphics[width=1\textwidth]{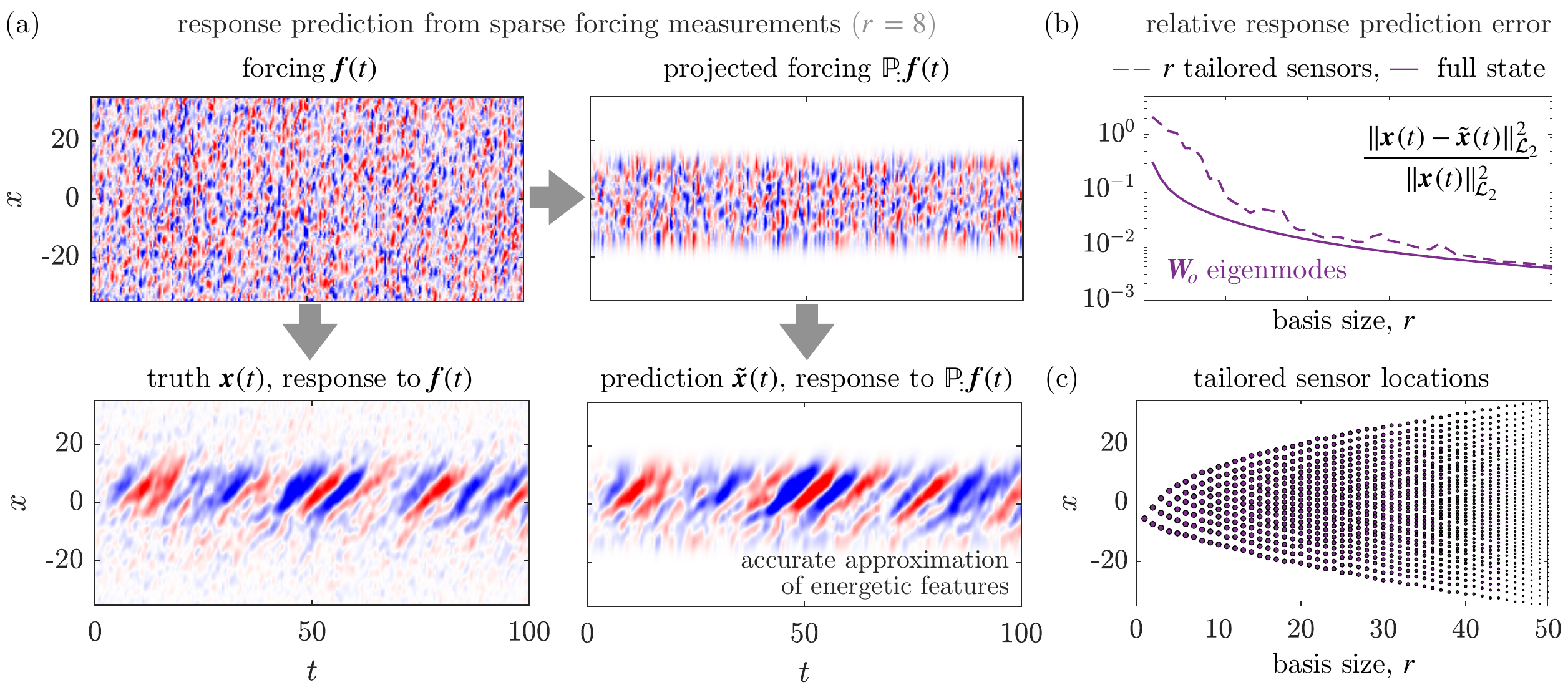}
    \caption{(a) Forcing applied to the linearized complex Ginzburg--Landau equation and its induced response compared to an interpolatory projection of the forcing and its induced response. The projector interpolates the forcing in the span of the leading $r=8$ eigenvectors of $\b{W_o}$ from measurements at $r=8$ spatial locations (sensors). (b) Normalized $\mathcal{L}_2$-error in the response prediction using interpolatory (dashed) and orthogonal (solid) forcing projections as a function of the projector rank $r$. (c) Sensor locations as a function of the size of the basis used for projection. Sampling points are selected using pivoted QR, as in \citet{manohar2018csm}, and are therefore tailored to the basis of $\b{W_o}$ eigenmodes. Plots show the real part of forcings and responses.}
    \label{fig:cgl2}
\end{figure}

\subsection{Sensor placement for prediction}

With a basis $\b{V}$ for the dominant forcing subspace for response prediction already identified, we now address how to interpolate the most responsive disturbances from sparse sensor measurements. As discussed in \S\ref{sec:methods}, this task reduces to the design of a sampling matrix $\b{P}$, containing columns of the identity, to be used in the construction of the interpolatory projector $\mathbb{P}_{:}=\b{V}\left(\b{P}^{\mathrm{T}} \b{V}\right)^{-1} \b{P}^{\mathrm{T}}$. To build a rank $r$ projector, we build $\b{P}$ using a pivoted QR factorization, as in \citet{manohar2018csm}, on the transpose of our basis containing the leading $r$ eigenvectors of $\b{W_o}$. The first $r$ QR-pivots correspond to our sampling points, or sensor locations, which we refer to as tailored sensors since they depend solely on $\b{V}$, and are therefore tailored specifically to that basis.\\

For our Ginzburg--Landau example, the most responsive disturbances buried within white noise forcing can be successfully exposed from sparse sensor measurements, as shown in Fig.~\ref{fig:cgl2}(a) for the case of $8$ sensors. The projected forcing captures the dynamically relevant parts of the disturbances, allowing for an accurate prediction of the energetic coherent structures in the response, as also shown in Fig.~\ref{fig:cgl2}(a). Moreover, the error induced by the interpolatory projection on the response prediction rapidly approaches the error induced by an orthogonal projector, built using the same basis, as the basis size increases, as shown in Fig.~\ref{fig:cgl2}(b). The spatial locations found for the sensors, tailored to the eigenvectors of $\b{W_o}$, closely resemble the optimal placements computed by \cite{chen2011jfm}. Furthermore, they also agree with the sensors locations found by \citet{manohar2021tac}, which is expected since those were tailored to the adjoint balancing modes, which span almost the same subspace as the eigenvectors of the observability Gramian.

\begin{figure}[t]
    \centering
    \includegraphics[width=1\textwidth]{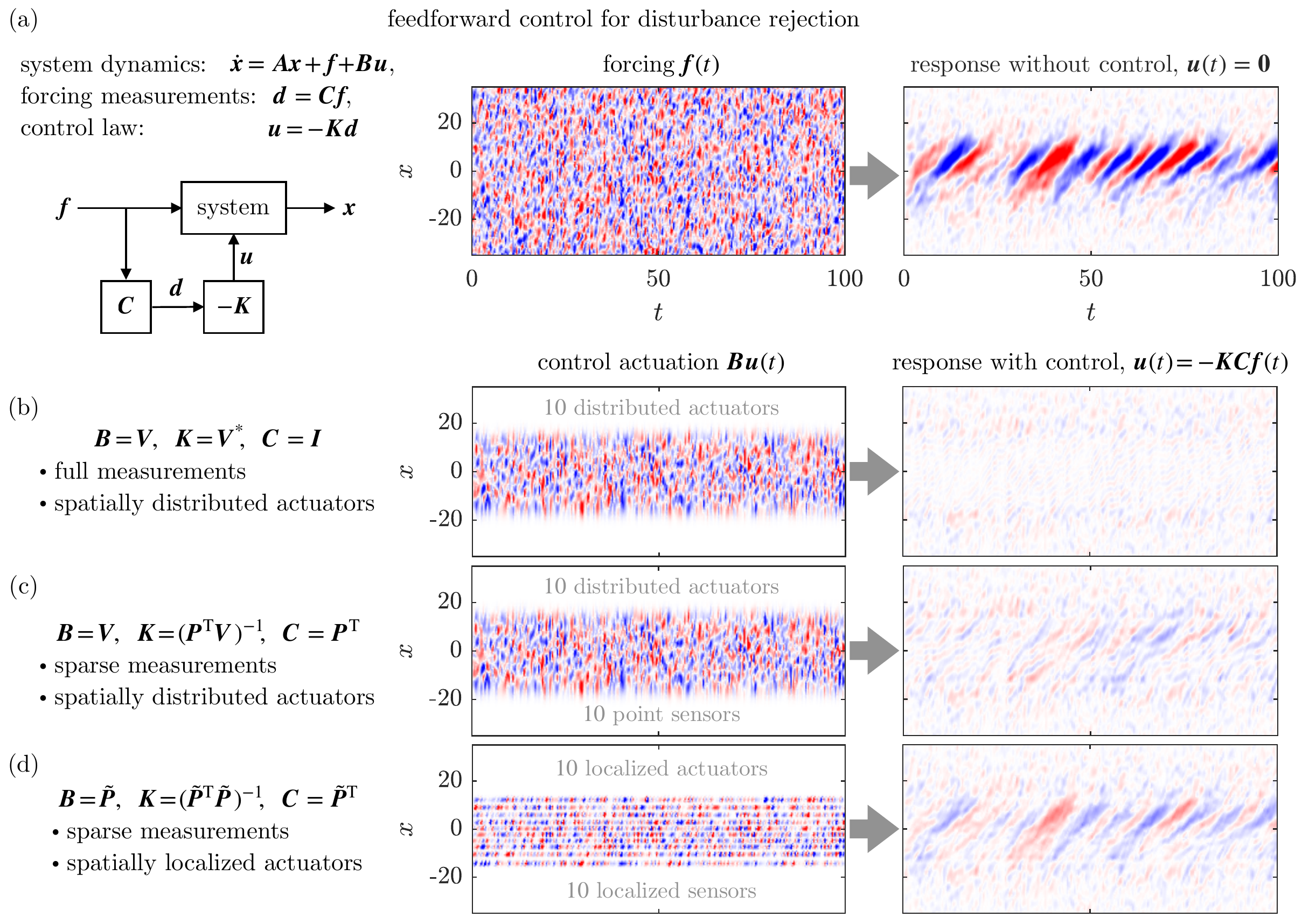}
    \caption{(a) Feedforward control strategy to reject disturbances in the linearized complex Ginzburg--Landau equation. The control actuation and the response of the controlled system is shown for three sensor and actuator configurations:(b) full disturbance measurements and $10$ spatially distributed (body forcing) actuators, (c) $10$ point sensors and $10$ spatially distributed actuators, and (d) $10$ localized sensors and actuators. Plots for the forcing, control actuations, and responses display the real part of these fields.}
    \label{fig:cgl3}
\end{figure}

\subsection{Feedforward control}

Consider a linear dynamical system
\begin{equation}
    \d{\b{x}}=\b{Ax}+\b{f}+\b{Bu}, \label{eq:ff}
\end{equation}
that is simultaneously forced by exogenous disturbances $\b{f}$ and the control actuation term $\b{Bu}$, with $\b{u}\in\mathbb{C}^p$ and $\b{B}\in\mathbb{C}^{n\times p}$. Disturbance feedforward control is an open-loop control strategy that uses measurements of the disturbances and a model of the dynamics to determine a sequence of control inputs with the objective of mitigating the effect of disturbances. Specifically, here we consider linear measurements of the disturbances $\b{d}=\b{Cf}$, where $\b{d}\in\mathbb{C}^q$ and $\b{C}\in\mathbb{C}^{q\times n}$, and control laws of the form $\b{u}=-\b{Kd}$, with $\b{K}\in\mathbb{C}^{p\times q}$, as shown in Fig.~\ref{fig:cgl3}(a). Substituting in Eq.~\eqref{eq:ff}, the controlled system dynamics become
\begin{equation}
    \d{\b{x}}=\b{Ax}+\tilde{\b{f}}, \qquad \tilde{\b{f}} = \left(\b{I}-\b{BKC}\right)\b{f},\label{eq:ff2}
\end{equation}
where the system is now forced by the effective forcing $\tilde{\b{f}}$.\\

We want to find a combination of actuators $\b{B}$, disturbance sensors $\b{C}$, and gain $\b{K}$ that minimizes the effect of $\tilde{\b{f}}$ on the response of the system. This can be achieved if $\left(\b{I}-\b{BKC}\right)$ projects $\b{f}$ onto the least responsive disturbances, or, equivalently, if $\b{BKC}$ projects onto the most responsive disturbances, given by the leading $r$ eigenvectors $\b{V}$ of the observability Gramian $\b{W_o}$. With this strategy in mind, we design three feedforward controllers for the linearized Ginzburg--Landau equation considering different restrictions on the allowable sensors and actuators.\\

First, we assume that we have access to full disturbance measurements, that is $\b{C}=\b{I}$, and that spatially distributed (body forcing) actuators are feasible. We design our actuators to directly excite the most responsive disturbances, that is $\b{B}=\b{V}$. Therefore, choosing the gain to be $\b{K}=\b{V}^*$, results in $\b{BKC}$ being an orthogonal projector onto the dominant forcing subspace. Considering $r=10$ distributed actuators, the feedforward controller is successfully able to reject disturbances by projecting-out the components that would cause the most energetic response, as shown in Fig.~\ref{fig:cgl3}(b). More precisely, a $97\%$ reduction in the $\mathcal{L}_2-$norm of the fluctuations is achieved.\\

For a second case, we allow only point-wise measurements of the disturbances, that is $\b{C}=\b{P}^{\mathrm{T}}$, and still consider spatially distributed actuators. Choosing our actuators to be $\b{B}=\b{V}$, means that the gain $\b{K}=\left(\b{P}^{\mathrm{T}}\b{V}\right)^{-1}$ makes $\b{BKC}$ an interpolatory projector. In this scenario, the full disturbance field is interpolated from sparse sensors, and then the actuators attempt to cancel out the components that would lead to the most energetic response. Using $r=10$ spatially distributed actuators and $10$ point sensors, selected via pivoted QR, the feedforward controller has only slightly lower performance than when using full measurements, as shown in Fig.~\ref{fig:cgl3}(c). Specifically, a $93\%$ reduction in the $\mathcal{L}_2-$norm of the fluctuations is obtained in this case.\\

Finally, we examine a more realistic scenario where only spatially localized sensors and actuators are allowed. Specifically, we consider $\b{B}=\tilde{\b{P}}$ and $\b{C}=\tilde{\b{P}}^{\mathrm{T}}$, where, rather than $r$ columns of the identity as in $\b{P}$, the columns of $\tilde{\b{P}}$ are given by discretized Gaussian functions
\begin{equation*}
    \exp\left( -\frac{\left(x-x_j \right)^2}{2\sigma^2}\right), \qquad j=1, \dots, r,
\end{equation*}
localized at sampling points $x_j$, where we choose $\sigma=0.6$. Using a control gain $\b{K}=\left(\tilde{\b{P}}^{\mathrm{T}}\tilde{\b{P}}\right)^{-1}$, the actuation then attempts to directly cancel out the disturbances measured in the regions localized around the sampling points.  We hypothesize that the optimal sensor locations for response prediction are also good for actuator placement, since they correspond to the optimal locations to interpolate the most responsive forcings. Therefore, leveraging the first $r=10$ eigenvectors of $\b{W_o}$, we use pivoted QR to select $10$ sampling points to place disturbance sensors and actuators. The resulting controller has a reasonable disturbance rejection performance, as shown in Fig.~\ref{fig:cgl3}(d). In this case, an $89\%$ reduction in the $\mathcal{L}_2-$norm of the fluctuations is achieved.

% comment regarding sparse forcing resolvent modes

\section{Leveraging response modes\label{sec:response}}

In this section, we present numerical examples detailing how the Gramian-based response modes can be leveraged to identify the most receptive perturbations of a dynamical system, and to reconstruct its state from sparse sensor measurements. As a test bed system we use a minimal channel flow at $\mathrm{Re}_{\tau}=185$. This corresponds to a pressure-driven turbulent flow, governed by the incompressible Navier--Stokes equations, in a doubly-periodic plane channel. The domain size is $1.83\times 2\times 0.92$ dimensionless length units along the $x$, $y$, and $z$ coordinates that indicate the streamwise (periodic), wall-normal, and spanwise (periodic) directions, respectively. For a Reynolds number of $\mathrm{Re}=4200$, based on the channel half-height $h$ and the centerline velocity for the laminar parabolic profile $U_{c,\mathrm{lam}}$, leading to $\mathrm{Re}_{\tau}=185$, this is the smallest domain that is able to sustain turbulence and is known as a minimal flow unit \citep{jimenez1991jfm}.\\

\begin{figure}[t]
    \centering
    \includegraphics[width=1\textwidth]{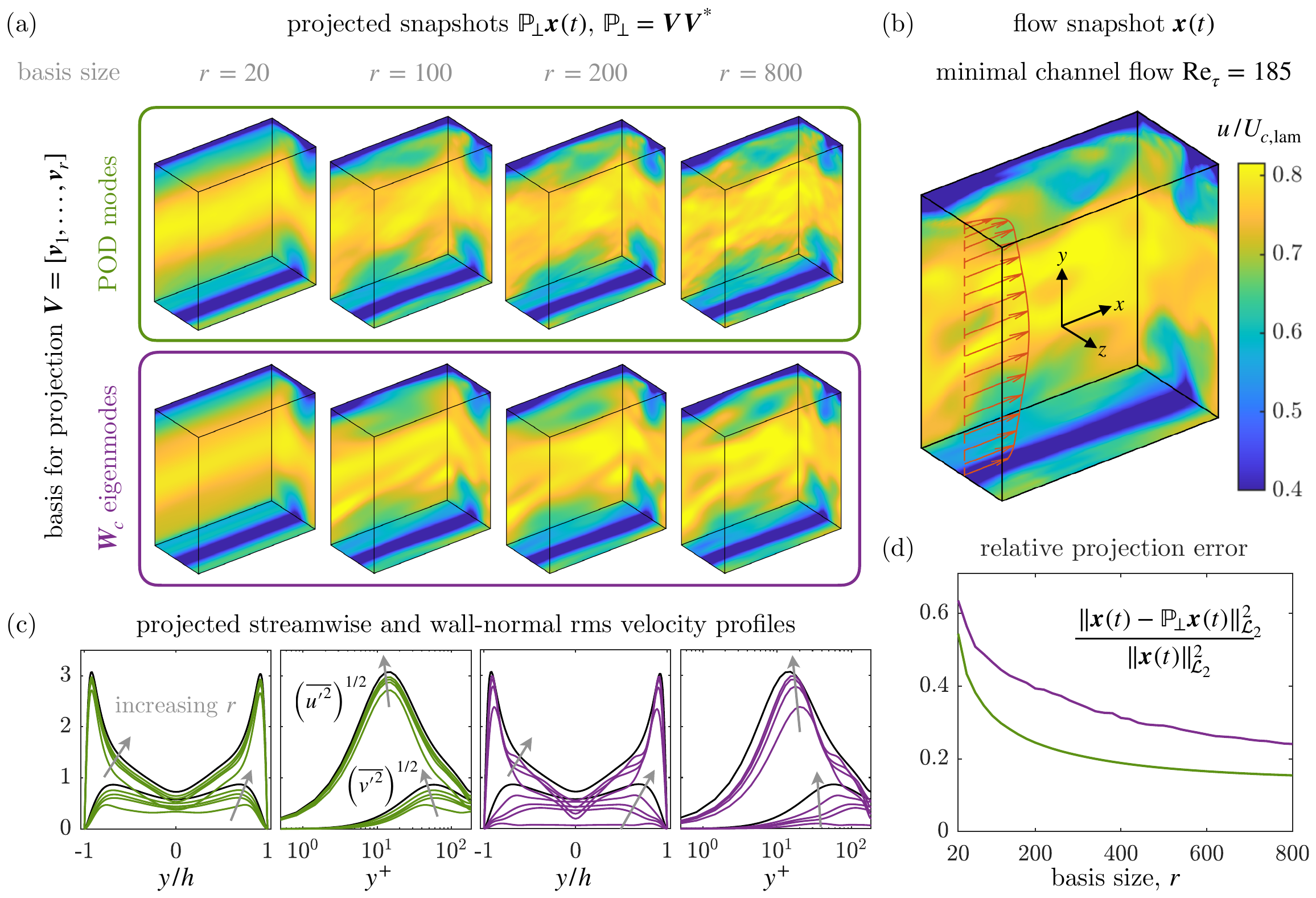}
    \caption{(a) Snapshots of minimal channel flow at $\mathrm{Re}_{\tau}=185$ projected onto subspaces spanned by the leading $r$ POD modes (top) and the leading $r$ eigenvectors of the controllability Gramian $\b{W_c}$ of the mean flow-linearized operator (bottom). (b) A typical DNS snapshot. The colormap shows streamwise velocity. (c) Streamwise and wall-normal rms velocity profiles of the projected snapshots at various projection ranks. Left panels (green curves) correspond to POD-based and right panels (purple curves) to Gramian-based projections. (d) Normalized $\mathcal{L}_2$-error in the response projections as a function of the projector rank $r$. POD modes span the optimal (in the $\mathcal{L}_2$-sense) linear subspace and therefore represent the lower bound for this error.}
    \label{fig:chflow1}
\end{figure}

\begin{figure}[t]
    \centering
    \includegraphics[width=0.92\textwidth]{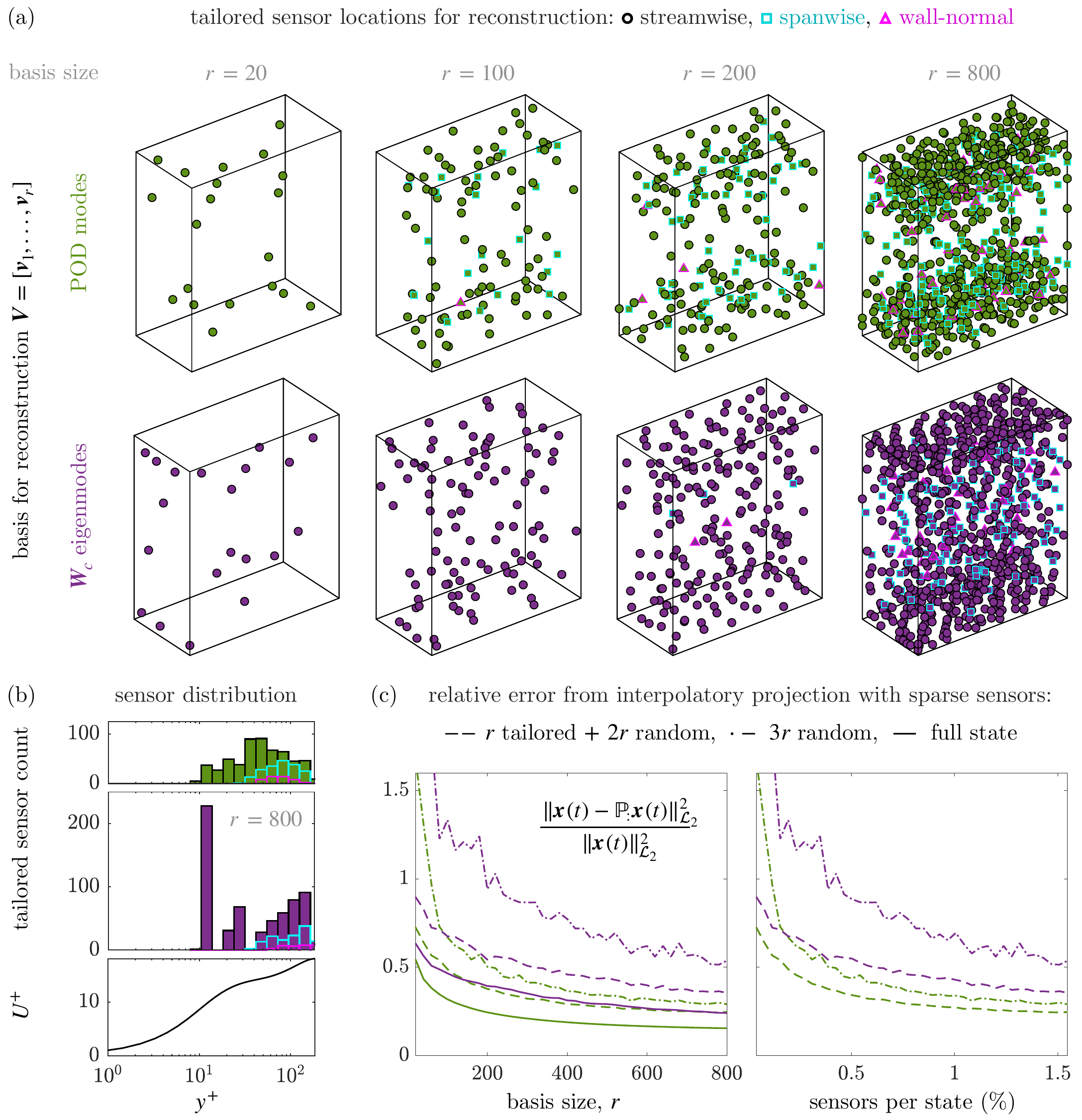}    \caption{(a) Spatial locations of sensors designed to reconstruct the velocity fluctuations in minimal channel flow at $\mathrm{Re}_{\tau}=185$. Sensors selected via pivoted QR \citep{manohar2018csm} are tailored to the leading $r$ POD modes (top) and the leading $r$ eigenvectors of the controllability Gramian $\b{W_c}$ (bottom). Sensors can measure streamwise (black circles), spanwise (cyan squares) or wall-normal (magenta triangles) velocity components. (b) Distribution of sensors in the wall-normal direction for $r=800$ using POD (top) and $\b{W_c}$ (bottom). (c) Normalized $\mathcal{L}_2$-error in the response projections as a function of the projector rank $r$ (left) and of the percentage of sensors over the total number of state variables (right). Three interpolatory projectors are considered for each basis using: $r$ tailored sensors complemented with $2r$ random sensors (dashed), $3r$ random sensors (dash-dotted), and all possible sensors (solid) which is equivalent to an orthogonal projector.}
    \label{fig:chflow2}
\end{figure}

We use the spectral code \emph{Channelflow} \citep{gibson2008jfm,gibson2014chflow} to perform direct numerical simulations (DNS). The code uses Chebyshev and Fourier expansions of the flow field in the wall-normal and horizontal directions, and a $3^{\mathrm{rd}}$-order Adams--Bashforth backward differentiation scheme for the time integration. We find that a grid with $32\times 101\times 16$ (after de-aliasing) in $x$, $y$, and $z$ and a time step of $0.005$ time units are sufficient to discretize the domain and keep the CFL number below $0.55$, for the cases studied. The flow is initialized from a random perturbation of the laminar profile, simulated for $10^4$ time units (based on $h/U_{c,\mathrm{lam}}$) until transients have died out and statistical stationarity is reached, and then velocity field snapshots are saved every $0.2$ time units for over an additional $1.5\times10^4$ time units, which is enough to get converged statistics. The streamwise velocity for a typical flow field snapshot is shown in Fig.~\ref{fig:chflow1}(b).\\

The mean flow is computed from the DNS snapshots and used to build the mean flow-linearized operator with an in-house code based on the Orr--Sommerfeld/Squire formulation. The code uses Chebyshev spectral collocation to discretize the wall-normal direction with the same grid used in the DNS. The linearized operator is built for every wavenumber tuple in the range resolved by the DNS and the controllability Gramian, considering full state inputs and outputs, is computed solving the corresponding Lyapunov equation with available routines. Eigenvectors of the global controllability Gramian $\b{W_c}$ are computed by taking the inverse Fourier transform of the eigenvectors obtained for every wavenumber combination and ordering them by their corresponding eigenvalues.\\

Orthogonal and interpolatory projections of velocity fluctuation snapshots from the DNS are investigated. The eigenvectors of $\b{W_c}$, arising from the mean flow-linearization, and POD modes, spanning the optimal linear subspace, are compared as bases for the projections. To compute the POD modes, the DNS dataset is augmented with reflections of the snapshots with respect to the midplanes in the $y$ and $z$ directions to enforce the corresponding symmetries. We then perform a randomized SVD \citep{halko2011siamrev} aiming at the leading $800$ POD modes using a Gaussian random matrix with $1000$ columns and without power iterations.

\subsection{Most receptive perturbations}

We examine the suitability of the eigenvectors of the controllability Gramian and POD modes as bases $\b{V}$ for the dominant response subspace. We build rank $r$ orthogonal projectors from the leading $r$ elements in these bases and use them to project the DNS snapshots. Qualitatively, both POD modes and $\b{W_c}$ eigenmodes seem to capture the main coarse features in the flow with a rank $100$ projection and further details as $r$ increases, as shown in Fig.~\ref{fig:chflow1}(a) for the streamwise velocity.\\

Second order statistics of the velocity fluctuations are approached by the those obtained for the projected snapshots as the rank of the projector increases with both bases, as shown in Fig.~\ref{fig:chflow1}(c). However, there is a slower convergence for the Gramian-based projection. Specifically, although the Gramian eigenmodes reasonably capture the peak in the rms profiles, a higher number of modes is required to properly resolve the statistics close to the wall and near the channel center.\\

More quantitatively, the response projection error, measured by the normalized $\mathcal{L}_2$-norm, drops faster with increasing $r$ and is always lower when using POD modes, as shown in Fig.~\ref{fig:chflow1}(d). This is expected because POD modes span the optimal linear subspace in the $\mathcal{L}_2$-sense and, therefore, represent the lower bound for the response projection error. Nevertheless, for this system, the error of the projection that leverages the eigenvectors of the controllability Gramian closely follow this lower bound. Importantly, this performance is achieved without any data snapshots and only requires knowledge of the mean flow, as opposed to the long sequence of high-fidelity snapshots needed to obtain converged POD modes. These results are not surprising since resolvent analysis has been very successful to model wall-bounded turbulent flows \citep{mckeon2017jfm} and, as we have shown, $\b{W_c}$ eigenmodes and the combination of all resolvent response modes at all frequencies span the same subspace.

\subsection{Sensor placement for reconstruction}

We now examine the performance of interpolatory projectors to reconstruct velocity fluctuations in the minimal channel flow from sparse sensor measurements. Again, sensor locations are selected using the pivoted QR approach from \citet{manohar2018csm}, and are thus tailored to the specific basis used. The sensor placement distribution reveals the spatial regions that are most informative to reconstruct the dominant coherent structures. Since the state vector of the system contains the three velocity components, sensors are also selecting which component to measure.\\

Leveraging both of our bases, of POD modes and of eigenvectors of $\b{W_c}$, we find that sensors accumulate towards the walls as more of them are added, as shown in Fig.~\ref{fig:chflow2}(a). A detailed view of the sensor distribution in the wall-normal direction, presented in Fig.~\ref{fig:chflow2}(b), reveals that most of the sensors tailored to the eigenbasis of $\b{W_c}$ are positioned closer to the wall and fall within the buffer region. For both bases, the great majority of the sensors measure streamwise velocity, which is expected since that is the most energetic component. Interestingly, there is more variety in the sensors tailored to the POD basis, including more measurements of the spanwise and wall-normal velocities, as shown in Fig.~\ref{fig:chflow2}(b).\\

A total of six projectors are built, three for each basis, and the reconstruction $\mathcal{L}_2$-error over the entire sequence of DNS snapshots is evaluated as a function of the projection rank $r$, as shown in Fig.~\ref{fig:chflow2}(c). 
Considering the leading $r$ elements in each basis, we build interpolatory projectors using $r$ tailored sensors complemented with an additional $2r$ random sensors. Because this is a high-dimensional system with a total of $n\approx 1.6\times 10^5$ states, complementing tailored sensors with random ones is a simple strategy to improve the resulting reconstruction that was suggested in the work of \citet{manohar2018csm}. This is implemented by simply taking the leading $2r$ columns of a matrix containing all remaining possible sensors after a random column permutation. These columns are then concatenated to the sampling matrix $\b{P}$ used to build the projector. For the sake of comparison, we also build interpolatory projectors using $3r$ random sensors and orthogonal projectors for the subspaces spanned by the POD modes and by the eigenvectors of $\b{W_c}$. Orthogonal projectors provide the lower bound in reconstruction error for interpolatory projectors onto the same subspace. Leveraging the controllability Gramian, the reconstruction using tailored sensors explain a large portion of the energy in the flow field, and, even with random sensors, $50\%$ of the energy is captured from measurements of $1.5\%$ of the state variables, as shown in Fig.~\ref{fig:chflow2}(c). Furthermore, this is achieved in a data-free manner, only requiring knowledge of the mean flow.

\section{\label{sec:conclusions}Conclusions}

In this work, we have drawn a connection between the resolvent operator and the Gramians from linear control theory. When considering full state inputs and outputs, the eigenvectors of the observability Gramian span the same subspace than the set of all resolvent forcing modes at all frequencies. The same is true for the eigenvectors of the controllability Gramian and the resolvent response modes. Moreover, we have shown that Gramian eigenmodes are the solutions to the optimization problems seeking for the most responsive forcing and receptive perturbation across all frequencies. Therefore, Gramian-based forcing and response modes form orthonormal bases of flow structures that are relevant over all frequencies, which makes them attractive for forcing and response projections in control-oriented applications. Through the examples presented, we have shown how to leverage interpolatory projections, built using readily available algorithms~\cite{manohar2018csm}, to place sensors and actuators for response prediction, feedforward control, and state reconstruction.\\

Eigenvectors of the observability Gramian provide an orthonormal basis of forcing modes that are hierarchically ordered by their responsivity. This means that, the response of a system that is being excited by high-dimensional disturbances can be approximated by the response to the projection of those disturbances onto the leading Gramian-based forcing modes. In fact, if all disturbances are equally likely, these modes provide the optimal forcing basis for response prediction. Although this is certainly not the case for turbulent flows, where there are preferred disturbances because nonlinearity provides colored forcing, we still expect these modes to provide an excellent, albeit sub-optimal, forcing basis for highly non-normal systems. This is supported by the bulk of resolvent literature for the case of wall-bounded turbulence~\citep{mckeon2017jfm}, since resolvent and Gramian-based forcing modes span the same subspaces. For instance,~\citet{bae2021jfm} showed that turbulence in a minimal channel flow is inhibited if the leading resolvent forcing mode (corresponding to the most dangerous frequency) is projected out of the nonlinear forcing at each time step of a numerical simulation. We hypothesize that the same would occur if the leading eigenvector of the observability Gramian were projected out instead, since both of these modes share almost identical spatial footprints. In the context of this work, this projection can be interpreted as feedforward control with full disturbance measurements (the full nonlinear forcing is known at each time step) and spatially distributed actuators (body forcing with the spatial footprint of the forcing mode).\\

Eigenvectors of the controllability Gramian provide an orthonormal basis of response modes that are hierarchically ordered by their receptivity. In other words, they provide a basis for the response of a forced system that is optimal, in the case of white-noise forcing, for state reconstruction. In fact, these modes coincide with POD modes for a linear stable system where every state is being disturbed. However, for turbulent flows, the eigenvectors of the mean-flow-linearized controllability Gramian form a sub-optimal basis for state reconstruction because of the colored forcing statistics. Nevertheless, for a high-dimensional system with disturbances applied everywhere, obtaining converged POD modes requires long data sequences, whereas the Gramian-based response modes require only knowledge of the base state (steady or mean flow). This is particularly appealing for the case of turbulent flows, where fully space- and time-resolved velocity field data snapshots can only be obtained numerically, whereas the mean flow can be obtained from experiments. Importantly, we showed that, for a minimal channel flow, the state reconstruction performance of the Gramian-based response modes follows closely that of POD, even though this is achieved without any data snapshots and only requiring knowledge the mean flow. Our findings also support the recent results by~\citet{cavalieri2022prf}, who built accurate and numerically stable reduced-order models of plane Couette flow via Galerkin projection onto controllability Gramian eigenmodes.\\

This work opens up several avenues of future research, including practical improvements of the proposed sensor and actuator placement strategy, its experimental application, and theoretical extensions to the connections drawn. For example, the presented method for sampling point selection, which leverages the Gramian eigenmodes and QR-pivoting~\citep{manohar2018csm}, is not able to account for the fact that, in experiments, some sensor and/or actuator configurations may be undesirable or even unfeasible. Therefore, penalizing or promoting certain sensor or actuator locations is of practical interest. This might be achieved using the bases of Gramian forcing and response modes in conjunction with the method developed by~\citet{clark2018sensors} to incorporate cost constraints. Another interesting research direction is to investigate the case of time-periodic flows, exploring if the connections presented extend to the harmonic resolvent~\cite{padovan2020jfm} and the frequential Gramians arising from the linearization about a periodic base flow~\cite{padovan2022arxiv}. The code developed for this work is available on github.com/ben-herrmann, and all the data is available for sharing upon request to the corresponding author.

\begin{acknowledgments}
We gratefully acknowledge Rahul Arun, Sean Symon, Samuel E. Otto, and Alberto Padovan for helpful comments and insightful discussions. This work was funded by ANID Fondecyt 11220465, the U.S. Office of Naval Research ONR N00014-17-1-3022, the Army Research Office ARO W911NF-17-1-0306, and the National Science Foundation AI Institute in Dynamic Systems 2112085.
\end{acknowledgments}

\appendix

\end{document}